# Hidden magnetic fields of the quiet Sun derived from Hanle depolarization of lines of the "second solar spectrum" at the limb from Pic du Midi observations

*Jean-Marie Malherbe (emeritus astronomer)*
Observatoire de Paris, PSL Research University, LIRA, France
Email: Jean-Marie.Malherbe@obspm.fr; ORCID: https://orcid.org/0000-0002-4180-3729
20 May 2025

**ABSTRACT**

This paper is based on a dataset of many strongly polarized solar lines belonging to the "second solar spectrum", i.e. the spectrum near the limb in linear scattering polarization. The observations were done at the Pic du Midi Turret Dome in 2006. The solar spectra were recorded at high spectral resolution (R = 400000) with the spectrograph slit orthogonal to the solar limb, so that $\mu = \cos\theta$ continuously varied from 0.0 to 0.45. The crystal liquid polarimeter delivered the linear polarization rate (Q/I). Strong lines such as CaII 3934 Å, CaI 4227 Å, SrI 4607 Å, SrII 4078 Å, BaII 4554 Å were studied. We measured the Hanle depolarization with the help of models predicting the polarization envelope with no magnetic field and we got values in the range 13-25 Gauss for the unresolved turbulent magnetic field, and we found that it often decreases towards the limb, revealing an altitude gradient. This present analysis was not yet published and spectra shown here become freely available to the research community.

**KEYWORDS**

Sun, polarization, spectra, high resolution, limb, Hanle depolarization, magnetic field, Calcium, Barium, Strontium

**INTRODUCTION**

The "second solar spectrum" is the linearly polarized spectrum observed at the limb. It is formed by coherent scattering due to the anisotropy of the radiation field. The linear polarization rate Q/I is generally small (< 1%) and decreases in the presence of magnetic fields (the Hanle effect). The first survey of the second solar spectrum was performed at Kitt Peak by Stenflo *et al.* (1983a, b). It was followed by a more sensitive survey by Gandorfer (2000, 2002) using the ZIMPOL polarimeter. The "second solar spectrum" requires polarization free telescopes to be best recorded and provides diagnostics of spatially unresolved, turbulent, and weak magnetic fields. Stenflo (2004) noticed from observations made at two different periods that the proportion of absorption-like and emission-like spectral lines changes, which suggests the varying influence of hidden magnetic fields with the cycle. The analysis of long-term fluctuations of the "second solar spectrum" could therefore be useful to monitor the magnetic cycle. Since the Pic du Midi Turret Dome is a polarization free refractor, it is convenient for this purpose.

**I - THE EXPERIMENT SETUP: SINGLE BEAM MODULATION FOR STOKES Q/I MEASUREMENTS**

The Turret Dome is a 50 cm polarization free refractor of 6.50 m focal length (Figure 1). However, a magnifying lens (X 5) provides an equivalent focal length of 33 m (F/66) at the secondary focus. The polarimeter of Figure 2 is located there, before injection into the spectrograph by two flat mirrors. It is made of Liquid Crystal Variable Retarders (LCVR) from the Meadowlark company (USA) followed by a linear polarizer, so that Stokes combinations I+Q, I-Q, I+U, I-U, I+V, I-V are got sequentially. For the "second solar spectrum", we used only one LCVR to observe I+Q and I-Q alternatively, as fast as possible. The polarimeter is followed by a 8 m Littrow type grating spectrograph (316 grooves/mm, 63°26' blaze angle) providing a typical dispersion of 5 mm/Å (Figure 2). The output spectrum (6 x 9 cm) is optically reduced 10 times and focused on an underline cooled CCD 12 bits (0 to 4095) camera (1370 x 1040 pixels) from LaVision (Germany) able to work with the LCVR at about 5 Hz cadence. The spectral pixel is 11 mÅ at 460 nm (X direction); the spatial pixel is 0.2" (Y direction). This cadence is not sufficient to freeze turbulence, but we do not look for spatial resolution; instead we are interested by polarimetric precision and the CCD accumulates 16 alternate couples of I+Q and I-Q and sums on line the 16 spectra of I+Q and I-Q. This elementary observation (2 x 16 exposures) is coded on 16 bits (0 to 65535) and is repeated many times during one hour (or more), so that we obtain finally hundreds or even thousands 16 bits FITS files. The data processing consists in correcting the line curvature and instrumental shifts or defaults (such as the fluctuating limb position, since we do not have any stabilization system), before summing together all the I+Q and I-Q spectra, resulting in 32 bits FITS files. The X direction of the FITS files is the wavelength while the Y direction is the abscissa along the slit, corresponding to the distance to the solar limb (1 pixel = 0.2"). The ultimate data are binned in the Y direction to have 1"/pixel. 2D maps of intensity and Q/I polarization rate are deduced and allow to draw the variation of Q/I at line centre or wings as a function of $\mu = \cos\theta$ (related to altitude by models). The full procedure is detailed by Malherbe *et al* (2007), but interpretations in terms of magnetic fields were unpublished.

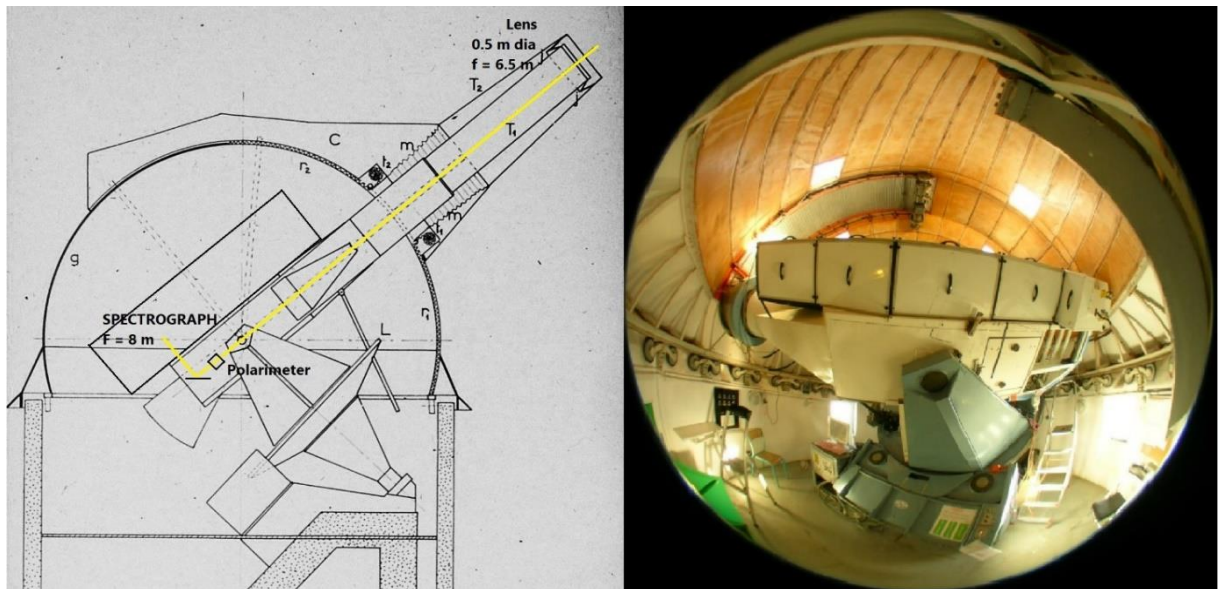

*Figure 1: the refractor of the Pic du Midi Turret Dome and attached 8 m spectrograph above*

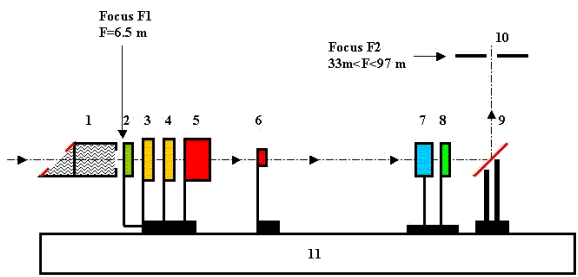 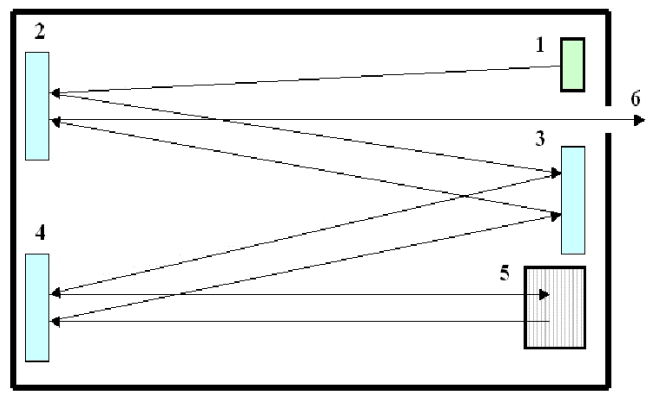

*Figure 2: the polarimeter (left) followed by the 8 m spectrograph (right)*

## II - RESULTS

CaI 4227 Å (Figure 3) exhibits high Q/I values at the limb (3.2 %), this is comparable to other determinations (3.6% at µ = 0.1 by Bianda *et al*, 1998; 2.6% in Gandorfer's atlas, 2000, 2002, also at µ = 0.1). At THEMIS we found (Malherbe, 2025) higher values (7% at µ = 0, 6% at µ = 0.1, 3.5% at µ = 0.15). Faurobert (1992) predicted possible values up to 8 % without any magnetic field. The observations seem to be depolarized by the Hanle effect in the presence of hidden magnetic fields.

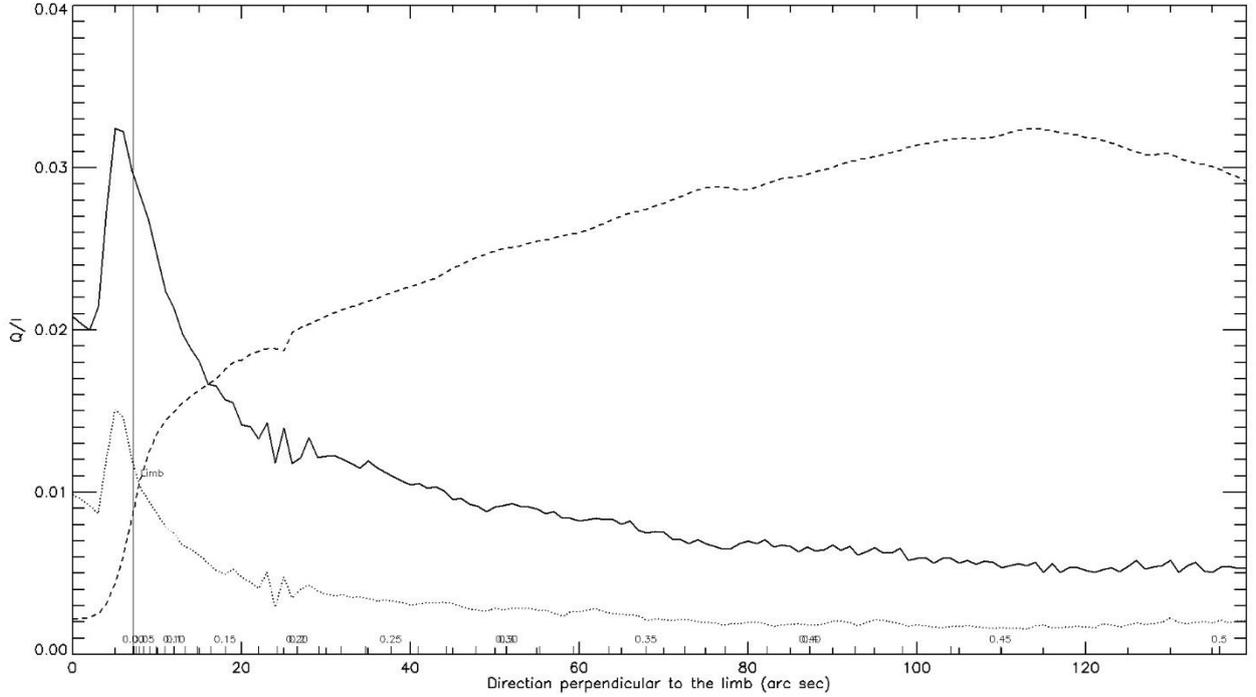

*Figure 3*: *polarization rate Q/I of the wing of CaI 4227 Å (solid line) and nearby continuum (dotted line) as a function of the limb distance (arcsec) or as a function of µ = cosθ. The chosen position of the limb (vertical bar) corresponds to the inflexion point of the line core intensity (dashed line).*

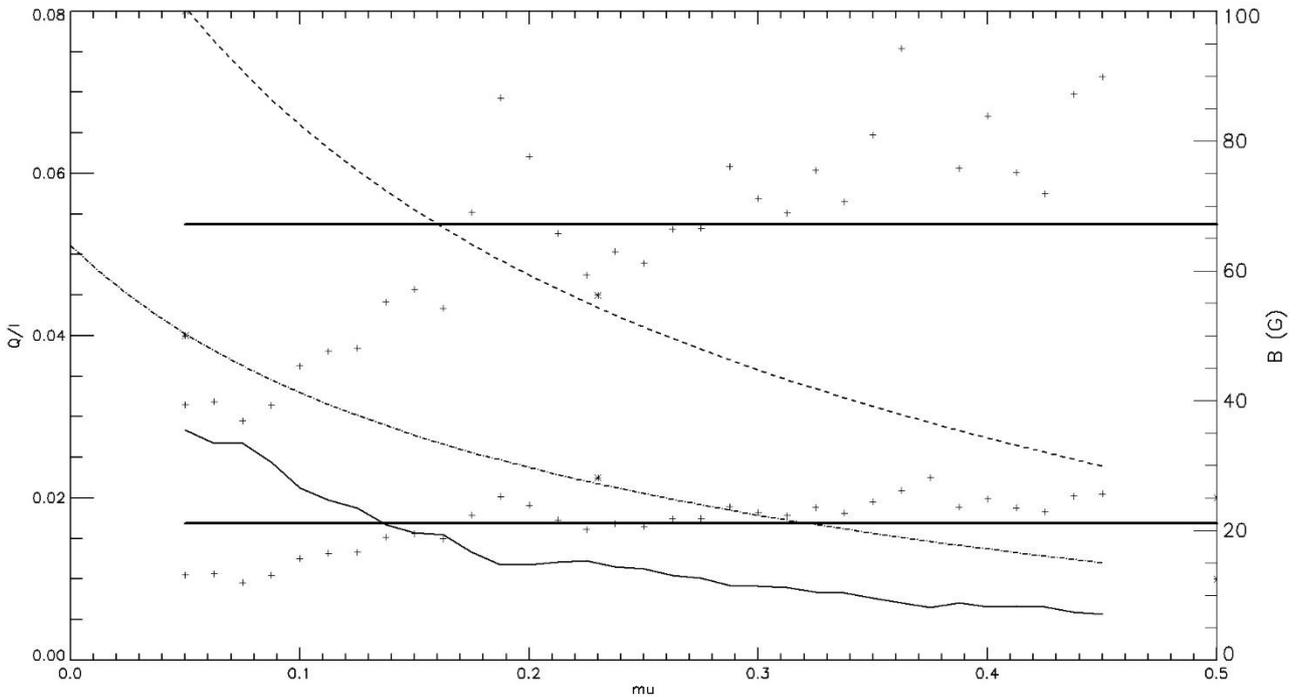

*Figure 4*: *observed polarization rate Q/I of the wing of CaI 4227 Å (solid line) and predicted maximum polarization (after Faurobert, 1992, two models) as a function of µ = cosθ (dashed lines). The crosses (+) represent the magnetic field strength; the horizontal bar is the mean value over µ.*

In order to derive from Figure 3 the magnetic field, we need to know the polarization with no magnetic field. We took the values computed by Faurobert (1992) for three values of µ (0.05, 0.23 and 0.5) and for other µ values, we used a fitting by the function $p(\mu) = a*(1 - \mu^2) / (\mu + b)$, a and b being constants computed by the least square method. We used two models, one for partial redistribution of frequencies, the second one being intermediate between total and partial redistribution (see Figure 3 of Faurobert's paper). This allows to get the Hanle depolarization in the wing of CaI. Then we used the Figure 4 of Stenflo (1982) to get the depolarizing magnetic field strength with the hypothesis of an isotropic distribution of the field vector. Stenflo's theory applies not only to CaI 4227 Å, but also to CaII 3934 Å, SrI 4607 Å, SrII 34078 Å and BaII 4554 Å, all these lines were observed by us at Pic du Midi. In case of partial redistribution, there is a strong depolarization so that we found B = 65 G in average over µ; in the intermediate case, the depolarization is much smaller and provides B = 20 G, µ averaged (Figure 4). We found also a clear behaviour for B to decrease towards the limb (with decreasing µ), corresponding to increasing altitude.

BaII 4554 Å (Figure 5) reveals Q/I values of 1.1% at µ = 0, this is quite similar to other studies (Malherbe, 2025, found 1.4% at µ = 0 at THEMIS; Stenflo & Keller, 1997, got 1.2% at µ = 0.1; 0.75% in Gandorfer's atlas, 2000, 2002, at µ = 0.1). We used the central peak (even isotope) which has 0 nuclear spin (so no hyperfine structure). We were not able to derive the magnetic field strength in the absence of predicted values of the line polarization with zero field.

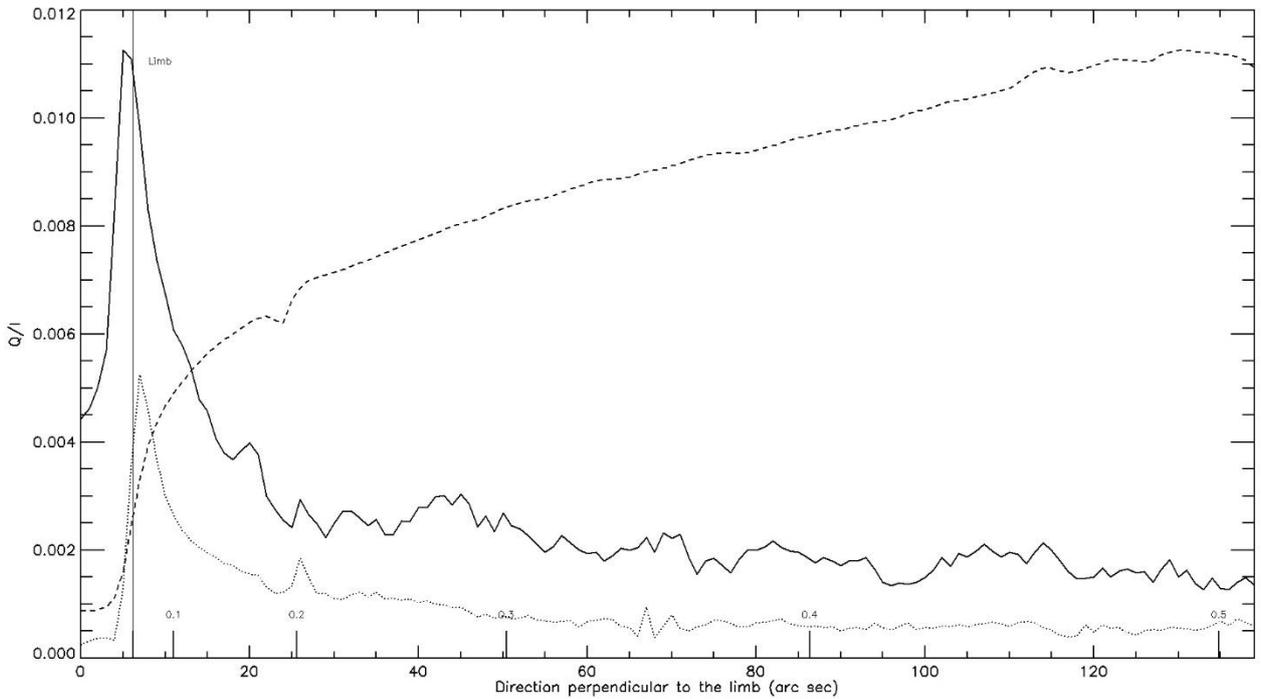

*Figure 5: polarization rate Q/I of the core of BaII 4554 Å (solid line) and nearby continuum (dotted line) as a function of the limb distance (arcsec) or as a function of µ = cosθ. The chosen position of the limb (vertical bar) corresponds to the inflexion point of the line core intensity (dashed line).*

SrI 4607 Å (Figure 6) reaches the maximum Q/I value of 1.8 %. This is close to other observations (2.1% at µ = 0 and 1.6% at at µ = 0.1 at THEMIS by Malherbe, 2025; 1.5% at µ = 0.1 by Stenflo & Keller, 1997; 1.2% in Bommier *et al*, 2005, and Gandorfer's atlas, 2000, 2002, at µ = 0.1). In order to derive from Figure 6 the magnetic field, we need the polarization free of magnetic field. We used the values computed by Faurobert (1993) for four values of µ (0.05, 0.1, 0.18 and 0.44) and for other µ values, we took a fitting by the function $p(\mu) = a*(1 - \mu^2) / (\mu + b)$. Then we used the Figure 4 of Stenflo (1982) to obtain the depolarizing magnetic field strength. We found (Figure 7) B = 13 G in average over µ. We did not find any clear behaviour for B to vary with altitude, contrarily to Bommier *et al* (2005) who suggested a gradient of -1 G for 10 km.

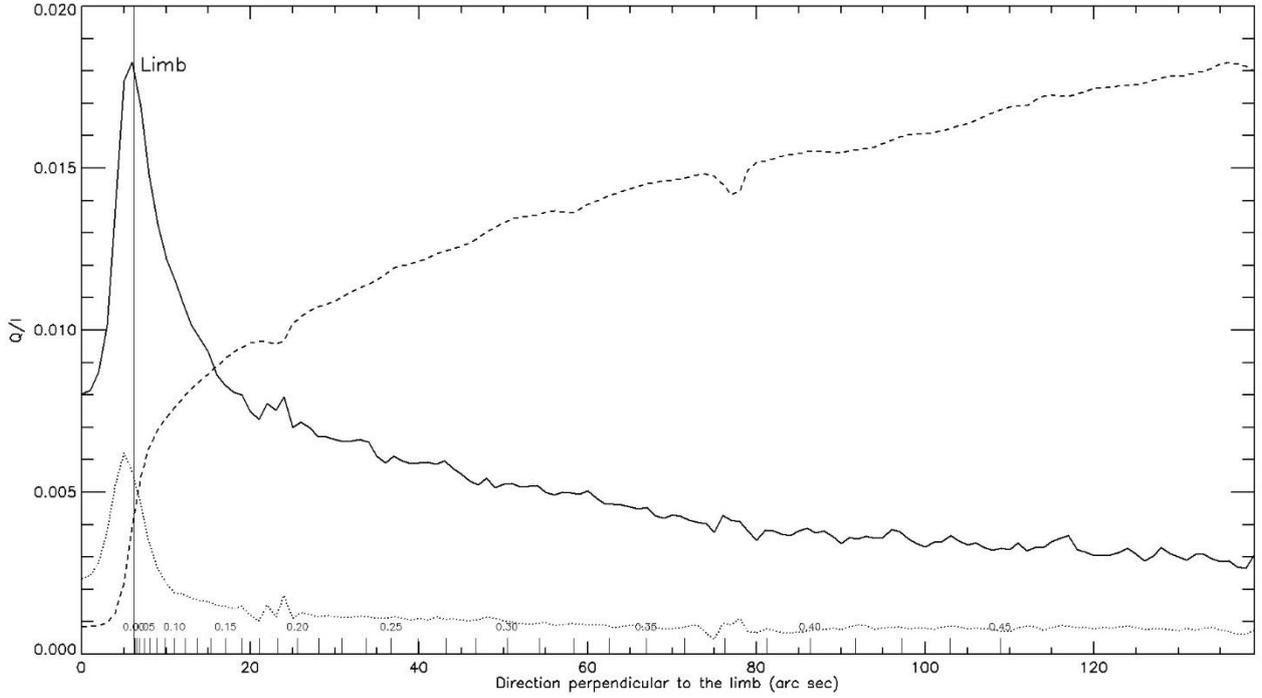

*Figure 6: polarization rate Q/I of the core of SrI 4607 Å (solid line) and nearby continuum (dotted line) as a function of the limb distance (arcsec) or as a function of µ = cosθ. The chosen position of the limb (vertical bar) corresponds to the inflexion point of the line core intensity (dashed line).*

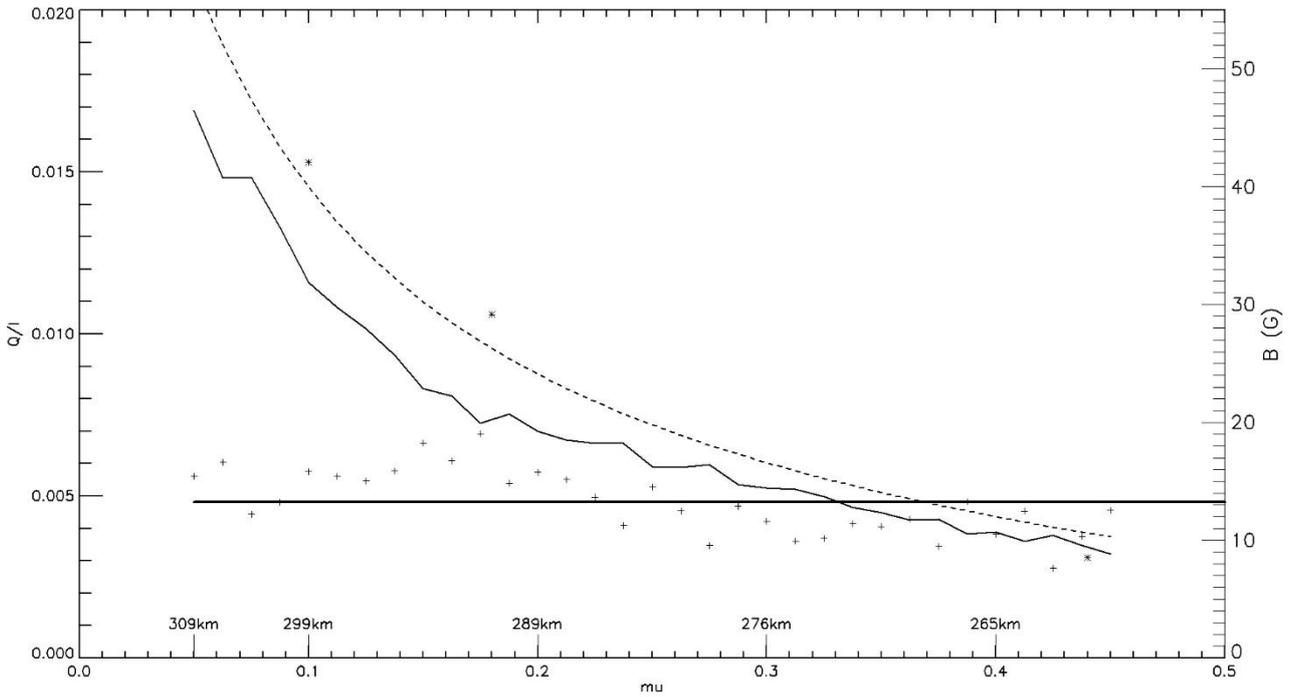

*Figure 7: observed polarization rate Q/I of the core of SrI 4607 Å (solid line) and predicted maximum polarization (after Faurobert, 1993) as a function of µ = cosθ (dashed line). The crosses represent the magnetic field strength; the horizontal bar is the mean value over µ. The altitude is after Bommier et al (2005).*

SrII 4078 Å (Figure 8) reveals Q/I values of 1.1% at µ = 0; this is close to the value of 1.5% at µ = 0.1 found by Stenflo & Keller (1997) and 1.1% of Gandorfer's atlas, 2000, 2002, also at µ = 0.1. In order to derive from Figure 8 the magnetic field, we need the polarization with no magnetic field. We used the values of Bianda (2003) for seven values of µ (0.05, 0.1, 0.15, 0.2, 0.3, 0.4 and 0.5) and for other µ values, we used the fitting by the function $p(µ) = a*(1 - µ²) / (µ + b)$. Then we found (Figure 9) B = 20 G in average over µ. As for CaI 4227 Å, Figure 9 shows clearly that B decreases towards the limb with decreasing µ (or decreases with altitude).

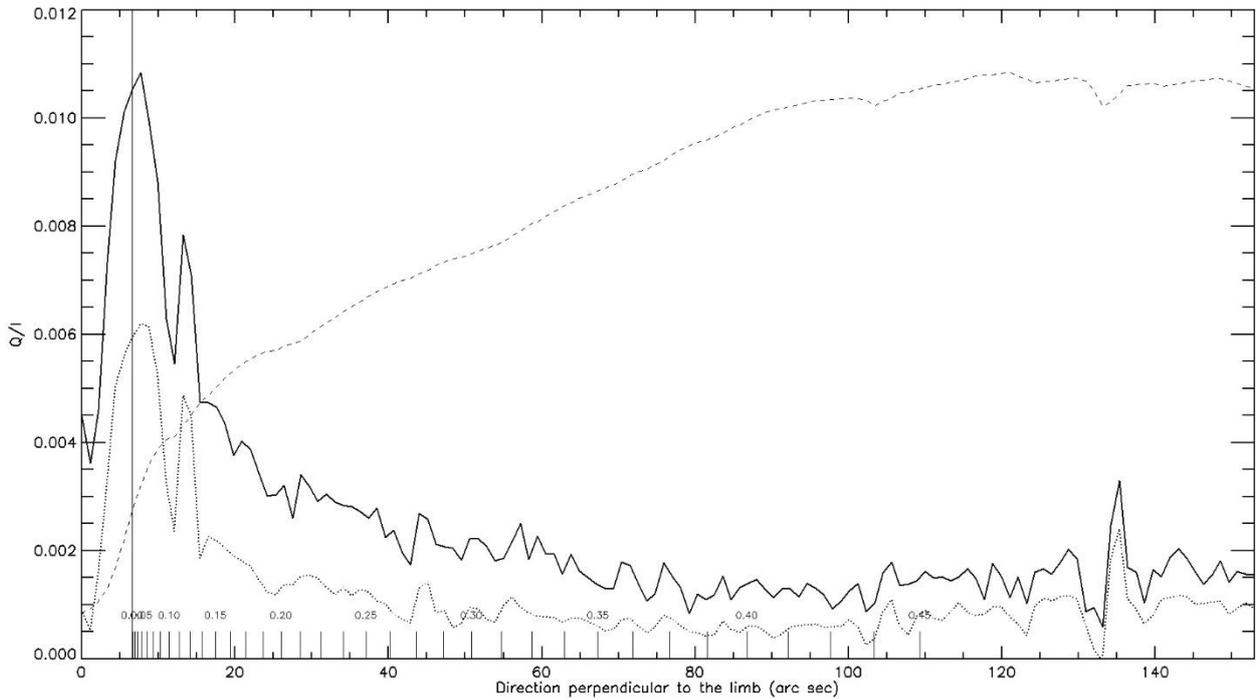

*Figure 8: polarization rate Q/I of the core of SrII 4078 Å (solid line) and nearby continuum (dotted line) as a function of the limb distance (arcsec) or as a function of µ = cosθ. The chosen position of the limb (vertical bar) corresponds to the inflexion point of the line core intensity (dashed line).*

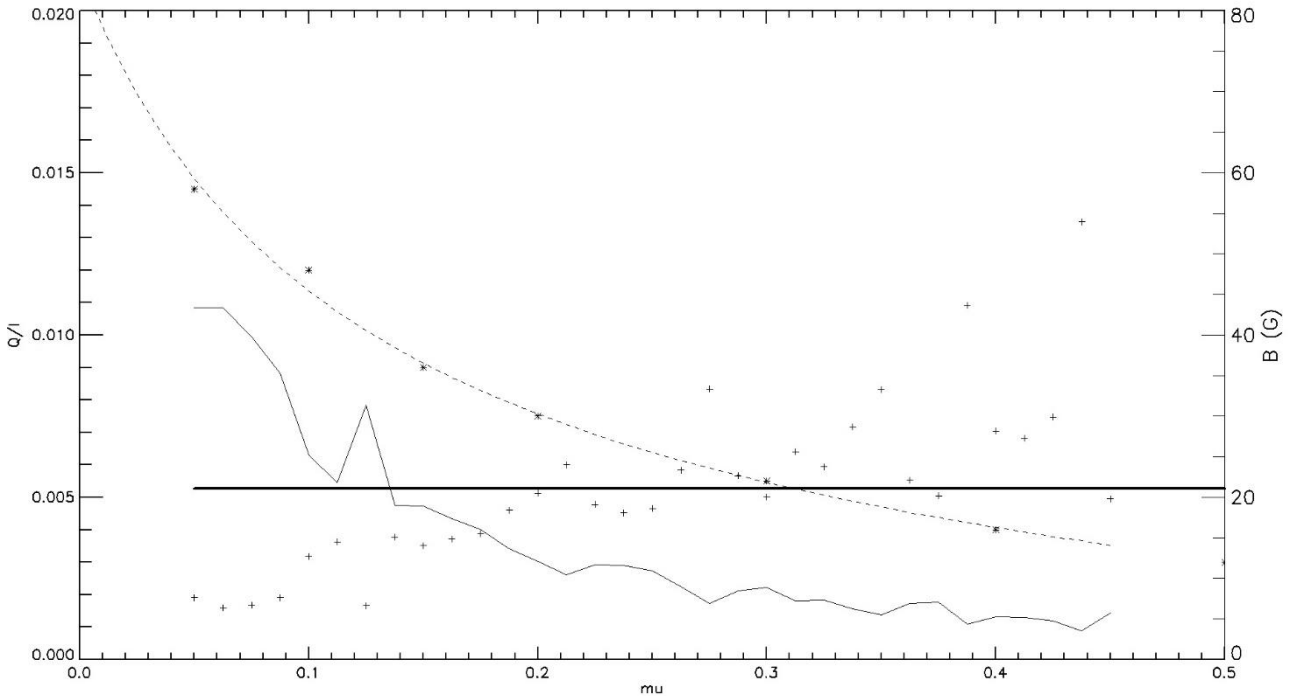

*Figure 9: observed polarization rate Q/I of the core of SrII 4078 Å (solid line) and predicted maximum polarization (after Bianda et al, 1993) as a function of µ = cosθ (dashed lines). The crosses represent the magnetic field strength; the horizontal bar is the mean value over µ.*

The situation was more complex for CaII K 3934 Å (Figures 10 & 11) because the line is chromospheric and presents many structures even at the limb. Observations were performed with the slit parallel to the limb for 5 values of µ (0.1, 0.15, 0.2, 0.3, 0.4), and also with the slit orthogonal to the limb. Our results are based on the 5 observations at constant µ, for which Figure 12 displays the polarization Q/I averaged along the slit at constant µ.

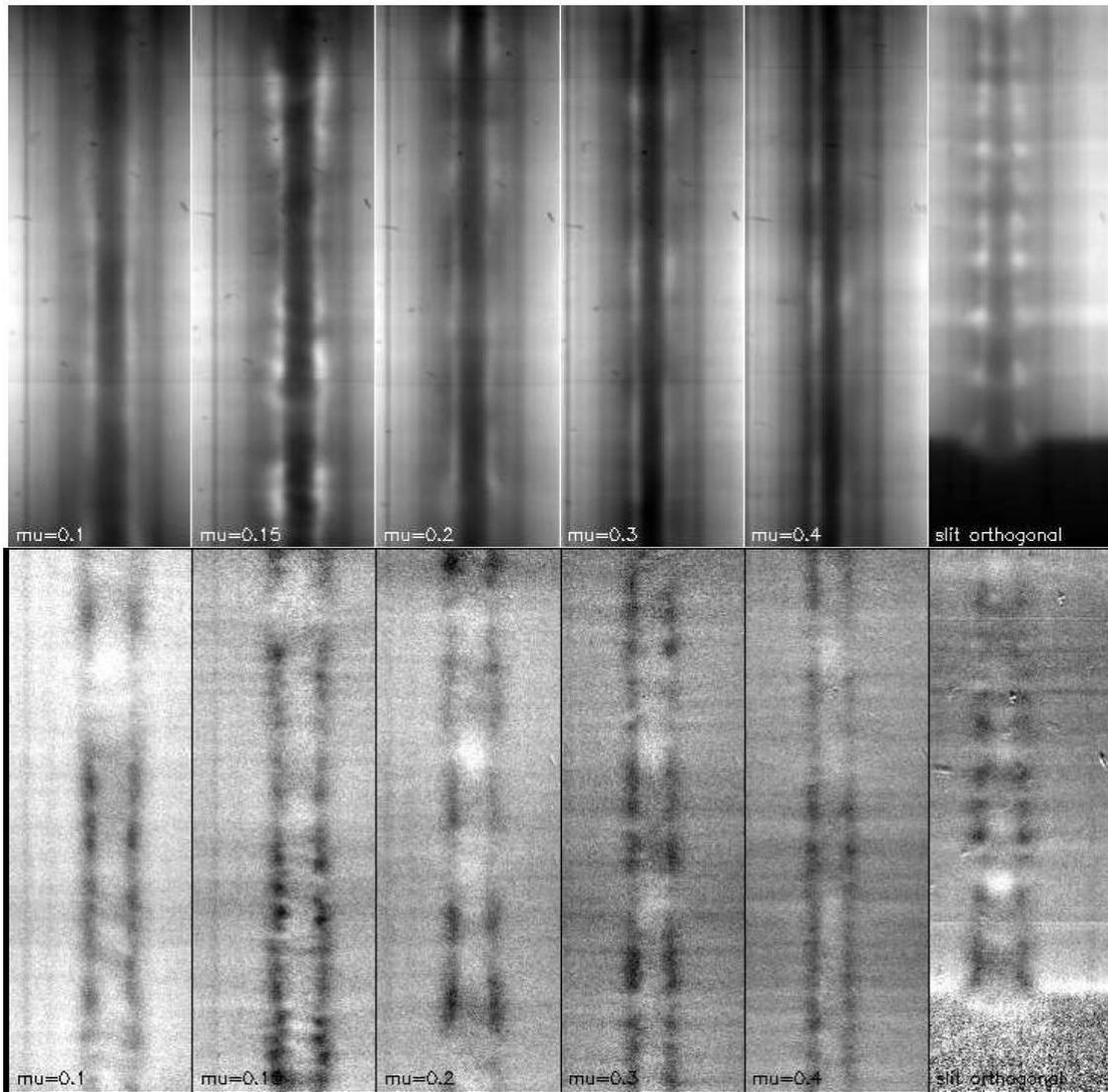

*Figure 10: Intensity (top) and polarization rate Q/I (bottom) of CaII K 3934 Å. For µ = 0.1, 0.15, 0.2, 0.3 and 0.4 the slit was parallel to the limb. In the right column, the slit was orthogonal to the limb.*

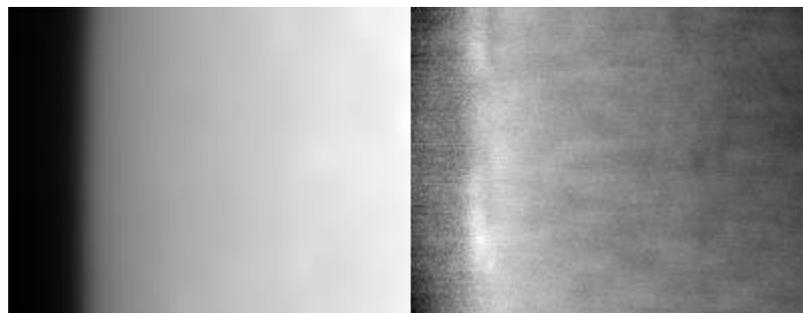

*Figure 11: observed intensity and polarization rate Q/I of the continuum at 393 nm through a broadband filter (10 nm FWHM).*

In order to derive from Figure 12 the magnetic field, we need the polarization with no magnetic field and used the values of Holzreuter & Stenflo (1993) for four values of µ (0.1, 0.2, 0.3, 0.4) and for other µ values, we used the fitting by the function $p(µ) = a*(1 - µ²) / (µ + b)$, a and b being constants. We found (Figure 13) B = 25 G for the isotropic depolarizing magnetic field strength in average over µ. As four atmospheric FAL models were used by Holzreuter & Stenflo (1993), there is a large error bar on the polarization envelope with no magnetic field; hence, we have also a large error bar on the field strength value, which inhibits to reveal clearly a possible atmospheric gradient.

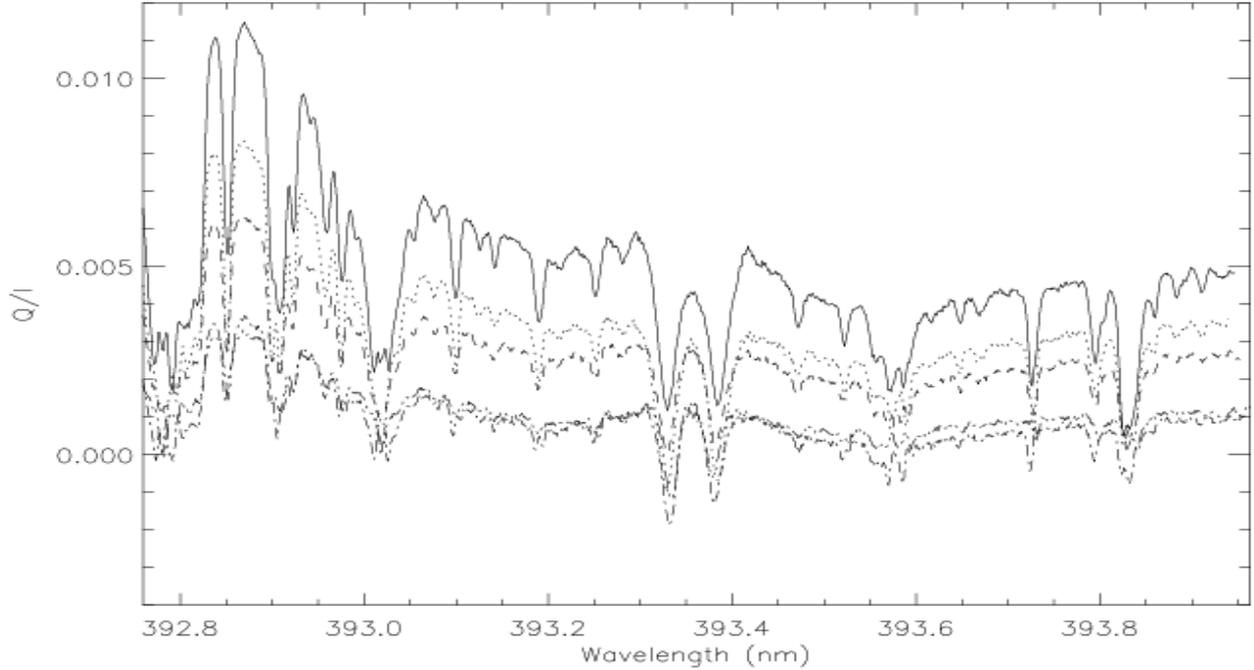

*Figure 12: polarization rate Q/I of CaII K 3934 Å for 5 values of µ (0.1, 0.15, 0.2, 0.3, 0.4)*

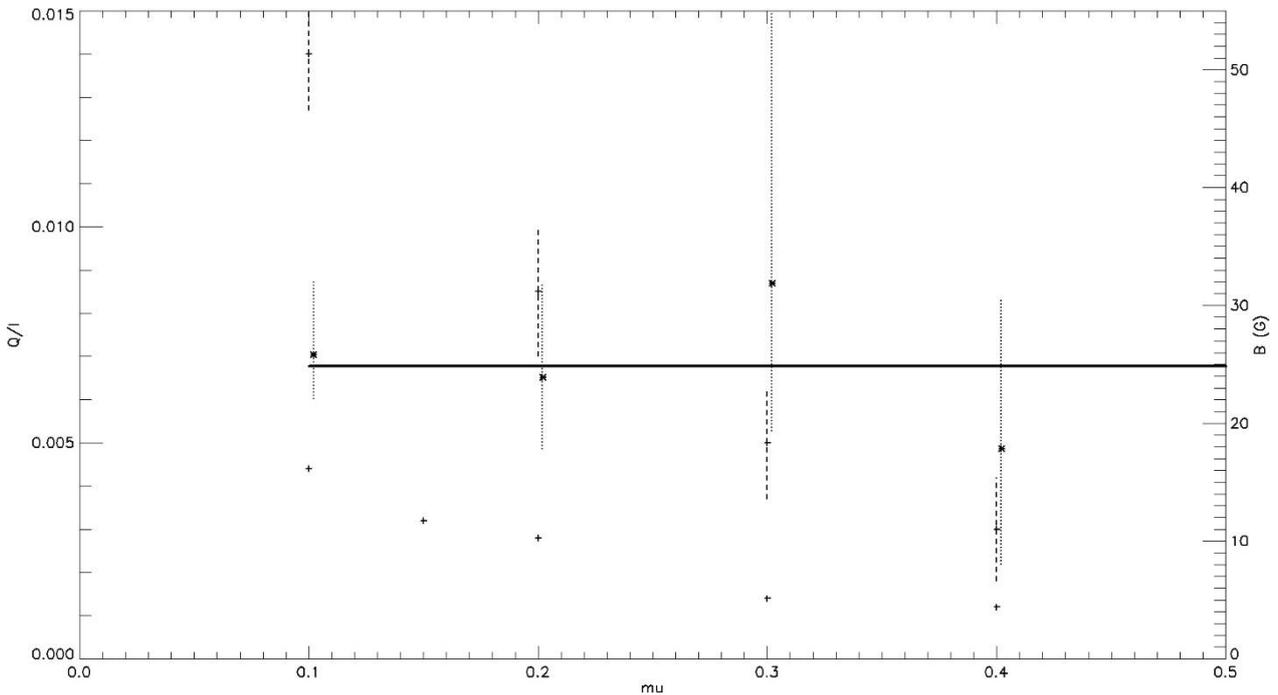

*Figure 13: observed polarization rate Q/I of the core of CaII K 3934 Å for 5 values of µ (0.1, 0.15, 0.2, 0.3, 0.4, symbols +), the envelope with no magnetic field from Holzreuter & Stenflo (dashed line = error bars) and field strength (symbols *, dotted line = error bar). The magnetic strength at µ = 0.3 may be due to some chromospheric structure. The values for µ = 0.1, 0.2 and 0.4 seem to indicate a decrease of the magnetic field strength towards the limb, as for CaI 4227 Å.*

## III - THE DATASET (ZIP ARCHIVE)

The dataset contains 5 directories of 32 bits FITS files corresponding to the 5 spectral lines, namely BaII, CaI, CaII, SrI and SrII.

The names of the 32 bits FITS files ("line" is the spectral line identification) are:
**int2D_line.fits** : 2D spectra of time averaged intensities $I(\lambda, y)$
**q2D_line.fits** : 2D spectra of time averaged polarization rate $Q/I(\lambda, y)$
(in x direction or wavelength, the step depends on the line, it is about 11 mÅ for SrI; in y direction or pixel position along the slit, we have 0.2"/pixel)
**int_cont_line.fits** : time averaged intensity of the continuum $I(y)$
**q_cont_line.fits** : time averaged polarization rate of the continuum $Q/I(y)$
**q_cent_line.fits** : time averaged polarization rate of the line core $Q/I(y)$

For CaII K line, we only have int2D_CaII.fits and q2D_CaII.fits files.

## CONCLUSION

Owing to the slit orthogonal to the solar limb, our measurements were performed as close as possible of the solar limb ($\mu \approx 0$) and in the vicinity of the limb ($\mu < 0.4$) for CaI 4227 Å, CaII 3934 Å, BaII 4554 Å, SrI 4607 Å and SrII 4078 Å. Hence, we have continuous curves of polarization $Q/I$ as a function of $\mu$. The Hanle depolarization allowed us to derive the turbulent magnetic field strength at various $\mu$ values. Most lines suggested that $\mu$ averaged strengths are in the range 13-25 G, according to the depolarization, and some lines indicated that there exists a negative field strength gradient (CaI, SrII), suggesting that the strength decreases with $\mu$ (or altitude) in the atmosphere. Such observations should be repeated many times along the solar cycle to study the fluctuations of the hidden, weak and turbulent, magnetic field.

**THE AUTHOR**

Dr Jean-Marie Malherbe (retired in 2023) is emeritus astronomer at Paris observatory. He first worked on solar filaments and prominences using multi-wavelength observations. He used the spectrographs of the Meudon Solar Tower, the Pic du Midi Turret Dome, the German Vacuum Tower Telescope, THEMIS (Tenerife) and developed polarimeters. He proposed models and MHD 2D numerical simulations for prominence formation. More recently, he worked on the quiet Sun, using satellites such as HINODE or IRIS, and MHD simulation results. He was responsible of the Meudon spectroheliograph from 1996 to 2023.